\documentclass[a4paper]{PoS}
\usepackage{amsmath, amssymb, bbm, physics}
\usepackage{setspace}
\usepackage{cleveref}
\usepackage{float}
\usepackage{diagbox}
\usepackage{comment}
\usepackage[stable]{footmisc}
\def\sup{\textsuperscript}

\title{$Sp(2N)$ Yang-Mills towards large $N$}

\ShortTitle{$Sp(2N)$ Yang-Mills towards large \textit{N}.}

\renewcommand*{\thefootnote}{\fnsymbol{footnote}}
\author{\speaker{Jack Holligan}\hspace{-.3em},\sup{\hspace{0.2em}$a$}\footnote{Supported by STFC Consolidated Grant ST/P00055X/1 and College of Science, Swansea University.}\hspace{0.4em} Ed Bennett,\sup{$b$}\footnote{Funded in part by the Supercomputing Wales project, which is part-funded by the European Regional Development Fund (ERDF) via Welsh Government.}\hspace{0.4em} Deog Ki Hong,\sup{$c$}\footnote{Supported by Basic Science Research Program through the National Research Foundation of Korea (NRF) funded by the Ministry of Education (NRF-2017R1D1A1B06033701).}\hspace{0.4em} Jong-Wan Lee,\sup{$c,d$}\footnote{Supported in part by the National Research Foundation of Korea grant funded by the Korea government (MSIT) (NRF-2018R1C1B3001379) and in part by Korea Research Fellowship programme funded by the Ministry of Science, ICT and Future Planning through the the National Research Foundation of Korea (2016H1D3A1909283).}\newline \hspace{0.4em} C.-J. David Lin,\sup{$e,f$}\footnote{Supported by the Taiwanese MoST grant 105-2628-M-009-003-MY4.}\hspace{0.3 em} Biagio Lucini,\sup{$g$}\footnote{Supported in part by the STFC Consolidated Grants ST/L00369/1 and ST/P00055X/1.}\hspace{0.7em}\footnote{Supported in part by the Royal Society Wolfson Research Merit Award WM170010.}\hspace{0.7em}\footnote{Received funding from the European Research Council (ERC) under the European Union's Horizon 2020 research and innovation programme under grant agreement No 813942.}\hspace{0.7em} Maurizio Piai,\sup{$a$}\footnotemark[7]\hspace{0.7em}\footnotemark[9]\hspace{0.7em} and Davide Vadacchino.\sup{$h$}\footnote{Acknowledges support from the INFN HPC-HTC project.}
\\
\llap{\sup{$a$}}Department of Physics, College of Science, Swansea University, Singleton Park, Swansea, SA2 8PP, UK\\
\llap{\sup{$b$}}Swansea Academy of Advanced Computing, Swansea University, Singleton Park, SA2 8PP, Swansea, UK\\
\llap{\sup{$c$}}Department of Physics, Pusan National University, Busan 46241, Korea\\
\llap{\sup{$d$}}Extreme Physics Institute, Pusan National University, Busan, 46241, Korea\\
\llap{\sup{$e$}}Institute of Physics, National Chiao-Tung University, Hsinchu 30010, Taiwan\\
\llap{\sup{$f$}}Centre for High Energy Physics, Chung-Yuan Christian University, Chung-Li 32032, Taiwan
\llap{\sup{$g$}}Department of Mathematics, Computational Foundry, Bay Campus, Swansea University, Swansea, SA1 8EN, UK\\
\llap{\sup{$h$}}INFN, Sezione di Pisa, Largo Pontecorvo 3, 56127 Pisa, Italy\\}
\abstract{Non-perturbative aspects of the physics of $Sp(2N)$ gauge theories are interesting for phenomenological and theoretical reasons, and little studied so far, particularly in the approach to the large-$N$ limit. We examine the spectrum of glueballs and the string tension of Yang-Mills theories based upon these groups. Glueball masses are calculated numerically with a variational method from Monte-Carlo generated lattice gauge configurations. After taking continuum limits for $N$ = 1, 2, 3 and 4, we extrapolate the results towards large $N$. We compare the resulting spectrum with that of $SU(N)$ gauge theories, both at finite $N$ and as $N$ approaches infinity.}

\FullConference{37\sup{th} International Symposium on Lattice Field Theory\\
                16-22 June, 2019\\
                Wuhan, Hubei Province, China}
\begin{document}

\renewcommand*{\thefootnote}{\arabic{footnote}}
\section{Introduction}
Gauge theories based on $Sp(2N)$ groups are of importance in beyond the Standard Model (BSM) physics. In particular, when matter consists of two Dirac fermions in the fundamental representation, the global $SU(4)$ flavour symmetry breaks down to $Sp(4)$. This symmetry breaking pattern gives rise to five pseudo-Nambu-Goldstone (pNGB) bosons. This mechanism provides a natural microscopic origin for the Higgs field as a composite particle, softening fine-tuning problems (see, for example, Refs.~\cite{Sp4,SP4NF2} and references therein for details). 

A relevant question for phenomenology is how observables depend on $N$. In order to get a first quantitative understanding of this dependence, here we investigate the large-$N$ limit of $Sp(2N)$ Yang-Mills. In particular, we compute the glueball spectrum of such theories for $N=1,3,4$; then, borrowing results for $N=2$ from Ref. \cite{Sp4}, an extrapolation to infinite $N$ is performed and comparisons are made with $SU(N)$ in the large-$N$ limit~\cite{Improved Smearing}. In addition, we investigate the Casimir scaling conjecture of~\cite{Casimir}. Our results will be reported in greater detail in \cite{SPNGlueballs}. 

\section{Lattice Model}
The symplectic group, $Sp(2N)$, can be defined in terms of the special unitary group of odd rank, $SU(2N)$, as follows:
\begin{equation}
    Sp(2N)=\{M\in SU(2N): M^*=\Omega^{\dagger}M\Omega\},\label{eq.Defn}
\end{equation}
where the symplectic matrix is conventionally defined to be
\begin{equation}\label{eq.Omega}
    \Omega = \left[\begin{array}{cc}
        0 & \mathbbm{1}_N \\
        -\mathbbm{1}_N & 0
    \end{array}\right],
\end{equation}
and $\mathbbm{1}_N$ is the $N\times N$ identity matrix. We use the notation $N_c = 2N$. Equation (\ref{eq.Defn}) is also the definition of pseudo-reality, given that $\Omega^{\dagger}=\Omega^{-1}$ from Eq. (\ref{eq.Omega}).

We regularise the $Sp(2N)$ Yang-Mills theory on a 4-dimensional Euclidean lattice of 4-volume $N_s^3\times N_t$ that is toroidal in all directions. The dynamics are governed by the Wilson action. If $g$ is the bare coupling and we denote a lattice link (which is an element of $Sp(2N)$) originating at site $x$ and in the direction $\hat{\mu}$ by $U_{\mu}(x)$, then the Wilson action is given by
\begin{equation}\label{eq.WilsonAction}
    S = \beta\sum_{x}\sum_{\mu<\nu}\left(1-\frac{1}{2N}\Re\Tr P_{\mu\nu}(x)\right),
\end{equation}
where the lattice coupling is defined as
\begin{equation}
    \beta=\frac{4N}{g^2}.
\end{equation}
In Eq.~(\ref{eq.WilsonAction}), $\Re$ and $\Tr$ denote real part and trace, respectively. $P_{\mu\nu}(x)$ is the plaquette, defined by
\begin{equation}\label{eq.plaquette}
    P_{\mu\nu}(x)=U_{\mu}(x)U_{\nu}(x+\hat{\mu})U^{\dagger}_{\mu}(x+\hat{\nu})U^{\dagger}_{\nu}(x).
\end{equation}
The four values of $\mu$ are 1, 2, 3 and 4 corresponding to the directions $x$, $y$, $z$ and Euclidean time $t$.

In this contribution, we are not considering the presence of dynamical fermions. The self-interaction of the gauge bosons gives rise to a spectrum of massive particles known as glueballs. We also study the physics of flux tubes (strings); if two infinitely massive static fermions are separated by a large distance, $r$, the potential between them is expected to behave as
\begin{equation}
    V(r)=\sigma r,
\end{equation}
for a confining theory, where $\sigma$ is the string tension, which can also be determined numerically.

Both the glueball spectrum and the string tension require a fully non-perturbative treatment of pure Yang-Mills and can be determined by a variational method applied to the lattice data.

\subsection{The Variational Method}\label{sec.Variational Method}
In the continuum, a glueball of spin $s$ exists in a $(2s+1)$-dimensional irreducible representation of the rotation group $SO(3)$. On the lattice, continuous rotational symmetry is broken to the octahedral group, $O_h$. The operators that create glueball states are gauge-invariant products of link variables that transform in the irreducible representation of $O_h$. Conventionally, the five irreducible representations of the cubic group are denoted by $R=A_1$, $A_2$, $E$, $T_1$ and $T_2$ (of dimensions 1, 1, 2, 3 and 3, respectively).\footnote{The labels for $T_1$ and $T_2$ are reversed compared to those used in Ref. \cite{Lucini:2010nv}.} The pseudo-reality of $Sp(2N)$ guarantees that charge conjugation is always positive and, as such, this quantum number is left implicit. The parity assignment $P$ of the state is denoted by a superscript ``$+$'' or ``$-$''. The general label for a particular state is $R^P$. Excited states are denoted by the addition of an asterisk (*) for each excitation. For example, $A_1^{+**}$ is the second excited state of the glueball in irreducible representation $A_1$ with positive parity and $A_1^+$ is the corresponding ground state.

We construct operators $\phi(\vec{x}, t)$ on the time slice \textit{t} stemming from spatial point $\vec{x}$ that are both gauge invariant and transform in a specific irreducible representation of the cubic group. The zero-momentum operator $\phi(t)$ is then obtained by summing $\phi(\vec{x}, t)$ over $\vec{x}$. Given $\phi(t)$, we construct the two possible parity eigenstates in the same channel. If $\hat{P}$ denotes the parity operator, the two states are made thus:
\begin{equation}
    \Phi^{\pm}(t) = \frac{1}{2}(\phi(t)\pm\hat{P}\phi(t)).
\end{equation}
We can determine the mass of a glueball in a given irreducible representation by computing the correlator of two such (vacuum-subtracted) operators: $\langle\Phi^{\dagger}(t)\Phi(0)\rangle$ where $\Phi(0)$ creates a state on the lattice at time 0 and $\Phi^{\dagger}(t)$ annihilates the state at time $t$. Without loss of generality, we label the energy eigenvalues in order of non-decreasing magnitude:
\begin{equation}
    E_1\leqslant E_2 \leqslant E_3 \leqslant \dots
\end{equation}
Inserting a complete set of glueball energy eigenstates into the correlator and making use of the time translation operator gives
\begin{eqnarray}
    \langle\Phi^{\dagger}(t)\Phi(0)\rangle &=& \sum^{\infty}_{n=1}\left|\langle n|\Phi(0)|0\rangle\right|^2e^{-tE_n}\equiv\sum^{\infty}_{n=1}|c_n|^2e^{-tE_n},
\end{eqnarray}
where the sum commences at $n=1$ since $\langle\Phi\rangle=0$. As $t$ increases, the lowest energy value will dominate the sum (assuming $|c_1|\neq 0$) and the correlator approximates a single exponential. Note that, having considered zero momentum operators, the energy $E_n$ coincides with the mass of the $n$-th excitation in the chosen channel. Below, the glueball mass of the ground state in the $R^P$ channel is denoted by $am(R^P)$, where $a$ is the lattice spacing.

The correlator used to measure the mass of the glueball decays exponentially, while background noise is a constant. Thus the statistical errors in the correlator become overwhelmingly large as the distance increases. This problem can be alleviated by increasing the value of $|c_1|$ in order to maximise the glueball signal at early times as well as decreasing the presence of excited states. To do this we construct a normalised $M\times M$ correlation matrix from a set of $M$ vacuum-subtracted operators $\{\Phi_i\}$ each transforming in the same irreducible representation of $O_h$:
\begin{equation}
    C_{ij}(t)=\frac{\langle\Phi^{\dagger}_i(t)\Phi_j(0)\rangle}{\langle\Phi^{\dagger}_i(0)\Phi_j(0)\rangle}.
\end{equation}
Asymptotically, eigenvectors of $C_{ij}(t)$ will decay as single exponentials with decay rates given by inverse masses of states in the considered channel. This calculation is explained in Refs.~\cite{Lucini:2010nv, Morningstar:1997ff}.

The string tension is measured in a similar fashion to the glueball spectrum, the difference being that the operator is the Polyakov loop which is defined at a spatial coordinate $\vec{x}$. It is computed by taking the trace of the time ordered product of lattice links in the Euclidean time direction:
\begin{equation}\label{eq.Polyakov}
    P(\vec{x})=\Tr\left[\prod^{N_t-1}_{t=0}U_4(\vec{x}, t)\right].
\end{equation}
As explained in section 5.1 of \cite{Sp4}, the correlator of two Polyakov loops separated in space allows us to determine the torelon mass from which we can extract the string tension in lattice units, denoted as $a^2\sigma$.

\subsection{Operators with a physical size}
Two additional tried and tested methods to improve the glueball signal at early times are blocking \cite{Blocking} and smearing \cite{Improved Smearing, Smearing}.\footnote{We refer the reader to the Refs. \cite{Improved Smearing, Smearing} for full details and formulae.} These methods can be combined with the use of multiple operators discussed in section \ref{sec.Variational Method}. We plan to determine the glueball masses in the continuum by extrapolating from finer and finer lattices. The operators themselves are the trace of a path ordered product of link variables on a fixed time slice, and the length of the closed path for each operator may be interpreted as a measure of size. The operators will decrease in size as the lattice spacing shrinks which raises two issues:
\begin{itemize}
\item The size of each operator will decrease while the glueball itself will retain a physical volume independent of the lattice spacing. This causes each operator to have a bad overlap onto physical states, hence decreasing the values of $|c_i|$.
\item The smaller lattice spacing will cause each link---and, hence, the operators---to be dominated by short-distance (UV) fluctuations. Consequently, the noise-to-signal ratio increases.
\end{itemize}

Smearing is an iterative process that serves to decrease UV fluctuations which become ever more pronounced at finer lattices. The algorithm adds to each link products of its neighbouring links weighted by some factor $p_a$. Like smearing, blocking is an iterative process but one that faster increases the physical size of the operators. At each step, the previous link is doubled in length and the neighbouring links are added with a weight factor $p_b$. 
The lattice links that undergo blocking are on a fixed time slice.
\subsection{Numerical Results}
\begin{figure}[ht]
    \centering
    \includegraphics[width=110mm]{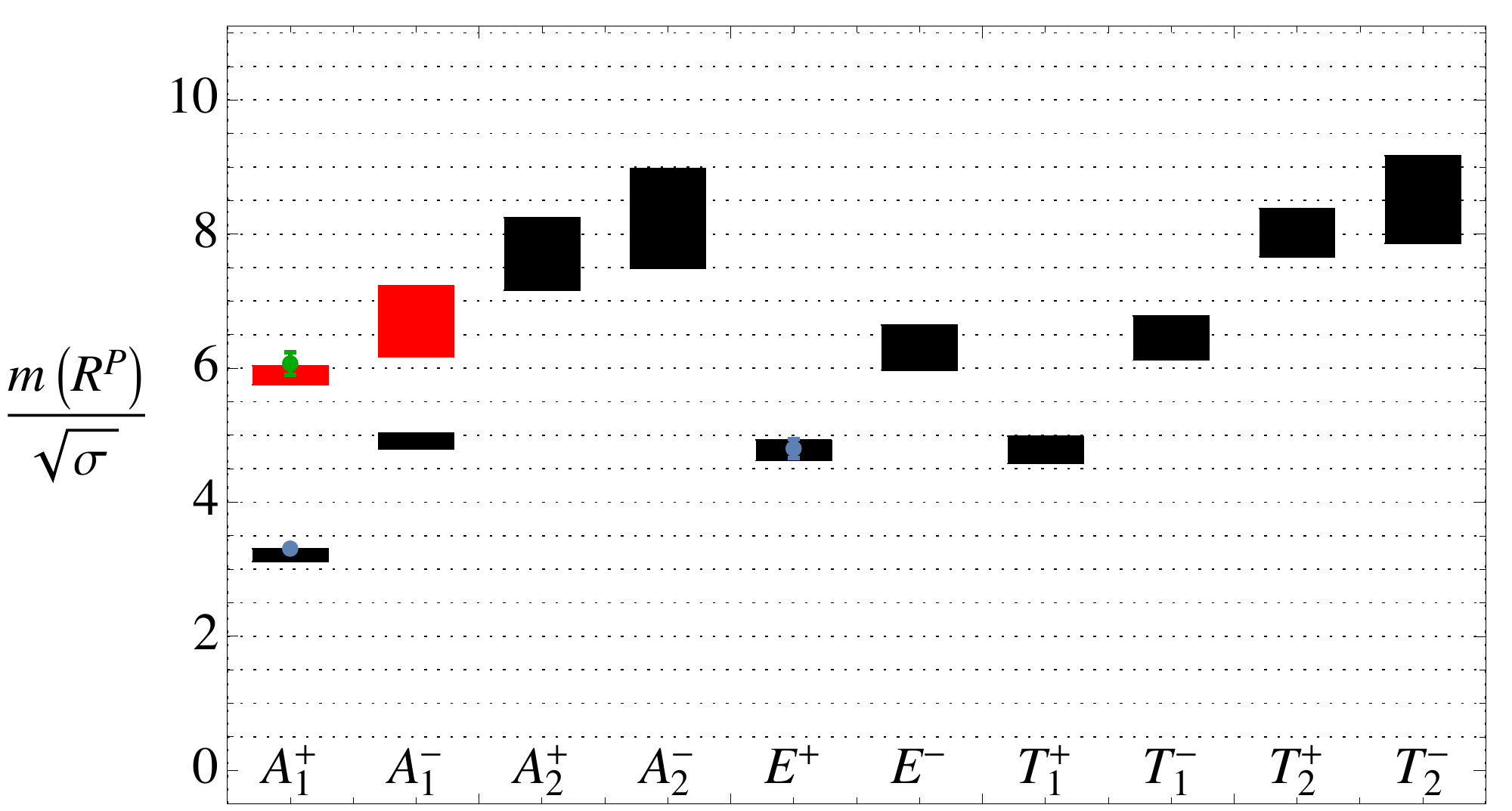}
    \caption{Result of large-$N$ extrapolation of the continuum glueball masses expressed in units of $\sqrt{\sigma}$.\footnotemark[3] Dots denote $SU(N)$ masses in the large-$N$ limit from Ref. \cite{Improved Smearing}. The blue ones denote ground states and the green ones denote first excitations. Boxes denote large-$N$ extrapolation of the continuum $Sp(2N)$ masses, their vertical thickness corresponding to the statistical error in the large-$N$ extrapolation. The latter receives contributions from both the error in the continuum extrapolations and from the direct measurement of the masses. The black boxes denote ground state masses and the red boxes denote first excitations. Glueball states are denoted by $R^P$, where $P = \pm$ is the parity assignment, while $R$ is the irreducible representation of the octahedral group.}
    \label{fig:Spectrum_MOverSigma}
\end{figure}

A single update in our case is defined by 4 over-relaxations and 1 heat-bath step to each of the links on the whole lattice. This is the same update process used in \cite{Sp4}, and we refer the Reader to that publication for further details. For a single value of $\beta$, we apply 10,000 thermalisation steps to equilibrate the lattice. After thermalisation, we save the lattice configuration every 10 updates to account for autocorrelation, until we have a total of 20,000 configurations. Using these configurations we compute the correlation matrices for each irreducible representation of $O_h$ described in section \ref{sec.Variational Method} to determine the glueball masses as well as the string tension. We repeat this for several $\beta$ values until we can perform a continuum extrapolation. We adopt an approximation, valid at leading order in $a$, defined as
\begin{equation}
    \frac{m(a)}{\sqrt{\sigma}}=\frac{m(0)}{\sqrt{\sigma}}+k_1a^2\sigma,
\end{equation}
where $k_1$ is an unknown constant to be determined empirically. 
After iterating for several values of $N$, we extrapolate to large $N$ with the ansatz
\begin{equation}
    \frac{m(2N)}{\sqrt{\sigma}}=\frac{m(\infty)}{\sqrt{\sigma}}+\frac{k_2}{2N},
\end{equation}
where $k_2$ is another unknown constant to be measured. Figure~\ref{fig:Spectrum_MOverSigma} reports the data of our extrapolation of the determined continuum $Sp(2N)$ glueball spectrum to infinite $N$ together with available continuum results for $SU(N)$ (taken from~\cite{Improved Smearing}). For states for which both are available, we note the expected agreement of the data for $SU(\infty)$ and $Sp(\infty)$.

\begin{figure}[ht]
    \centering
    \includegraphics[width=110mm]{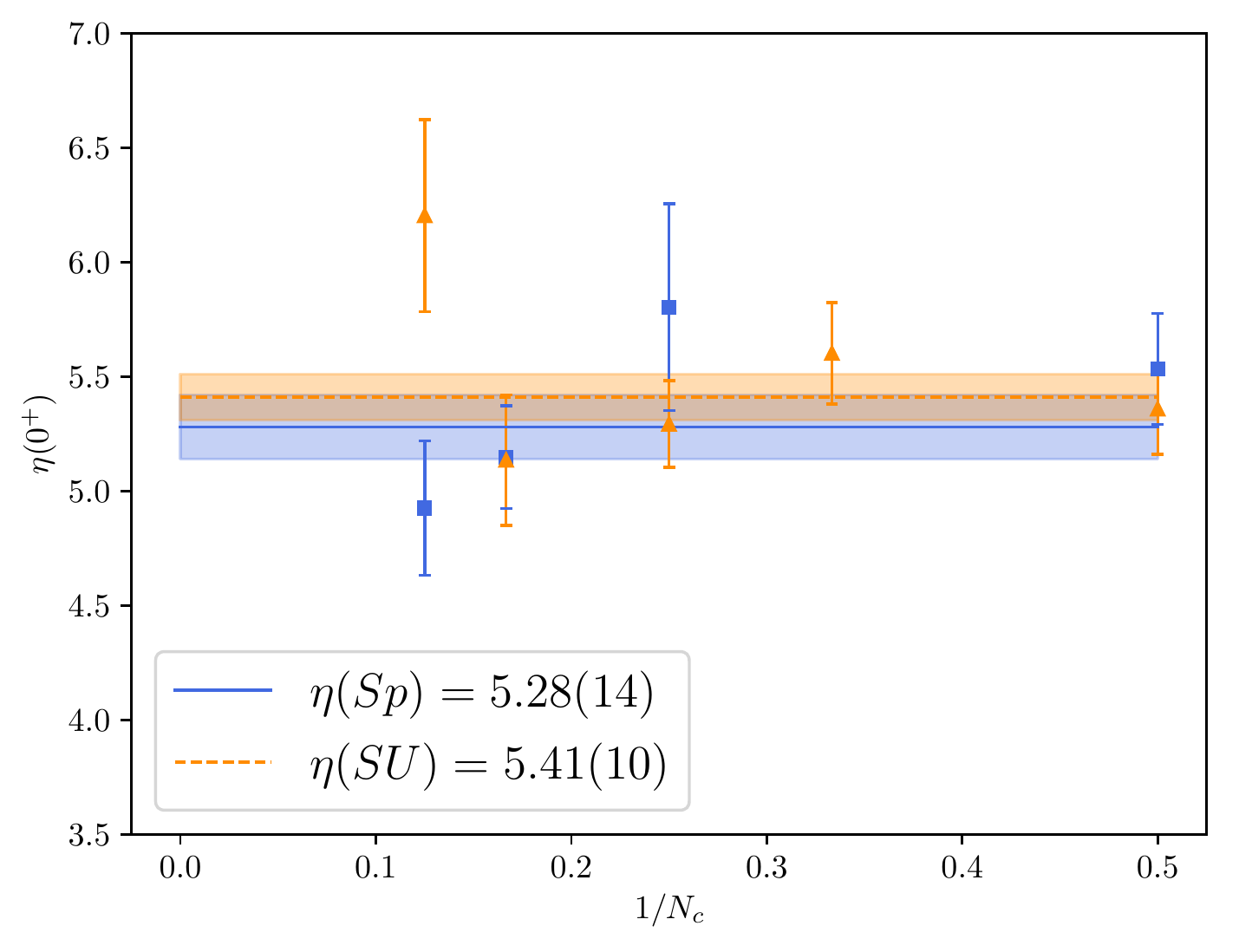}
    \caption{The quantity $\eta(0^+)$ defined in Eq. (\ref{eq.Casimir Scaling}) expressed as a function of $1/N_c$.\footnotemark[4] The orange points correspond to $SU(N_c)$ data ($\blacktriangle$) and the vertical width corresponds to the error. The blue points correspond to $Sp(N_c)$ data ($\blacksquare$) and the vertical width corresponds to the error. The two strips are the values of $\eta(0^+)$ for the group of matching colour with vertical width corresponding to the error.}\label{fig:Casimir}
\end{figure}

In addition to comparing large-\textit{N} glueball masses between the groups, we can also test on our data the conjecture of Casimir scaling described in \cite{Casimir}. We can define the ratio
\begin{equation}\label{eq.Casimir Scaling}
    \eta(0^{+})\equiv\frac{m^2_{0^{+}}}{\sigma}\frac{C_2(F)}{C_2(A)}.\end{equation}
$C_2(F)$ and $C_2(A)$ denote the quadratic Casimirs for, respectively, the fundamental and adjoint representations of the gauge group. We restrict our attention to glueball states with positive charge conjugation. The ratios of $C_2(F)$ and $C_2(A)$ are given in Eq. (4) of Ref. \cite{Casimir}. The conjecture is that $\eta$ is a constant that depends only on the number of spacetime dimensions. As show in Fig. \ref{fig:Casimir} and Tab. \ref{tab.Casimir}, Casimir scaling is supported by our calculation.

\begin{table}
\centering
    \begin{tabular}{|c|c|c|}
    \hline
    Group & $\eta(0^{+})$ & $\chi^2/\text{N\textsubscript{d.o.f.}}$\\
    \hline
    \textit{$SU(N_c)$}:&5.41(10)&1.43\\
    \textit{$Sp(N_c)$}:&5.28(14)&1.42\\
    \hline
\end{tabular}
\caption{The value of $\eta(0^+)$ in Eq. (\ref{eq.Casimir Scaling}) computed from $SU(N_c)$ data and $Sp(N_c)$ data.\footnotemark[3] The rightmost column is the reduced chi-squared for the data fitted to the value of $\eta$. All of the first row is quoted from Ref. \cite{Casimir}.}
\label{tab.Casimir}
\end{table}
\section{Conclusion}
Within this contribution we have considered pure $Sp(2N)$ Yang-Mills theories on the lattice. We computed the glueball spectrum for such theories at various $N$ and extrapolated $N$ to infinity. We found agreement with the glueball spectrum of $SU(N)$ groups in the same limit and found evidence to support Casimir scaling.
\footnotetext[3]{Updated since presented at \textit{Lattice 2019} by making use of larger statistics.}
\section{Acknowledgements}
JH wishes to thank all co-authors for their guidance throughout this project as well as David Schaich for his follow-up question and discussion on the chemical potential in $Sp(2N)$.
Numerical simulations have been performed on the Swansea SUNBIRD system, on the local HPC clusters in Pusan National University (PNU) and in National Chiao-Tung University (NCTU), and on the Cambridge Service for Data Driven Discovery (CSD3). The Swansea SUNBIRD system is part of the Supercomputing Wales project, which is part-funded by the European Regional Development Fund (ERDF) via Welsh Government. CSD3 is operated in part by the University of Cambridge Research Computing on behalf of the STFC DiRAC HPC Facility (www.dirac.ac.uk). The DiRAC component of CSD3 was funded by BEIS capital funding via STFC capital grants ST/P002307/1 and ST/R002452/1 and STFC operations grant ST/R00689X/1. DiRAC is part of the National e-Infrastructure.

\footnotetext[4]{Updated since presented at \textit{Lattice 2019} by making use of larger statistics.}
\end{document}